\documentclass[conference]{IEEEtran}

\IEEEoverridecommandlockouts
\usepackage{cite}
\usepackage{amsmath,amssymb,amsfonts}
\usepackage{algorithmic}
\usepackage{graphicx}
\usepackage{textcomp}
\usepackage{xcolor}
\def\BibTeX{{\rm B\kern-.05em{\sc i\kern-.025em b}\kern-.08em
    T\kern-.1667em\lower.7ex\hbox{E}\kern-.125emX}}

\usepackage[english]{babel}

\usepackage{amsmath}
\usepackage{graphicx}
\usepackage[colorlinks=true,  allcolors=blue]{hyperref}

\usepackage{cite}
\usepackage{matlab-prettifier}
\usepackage{cleveref}
\usepackage{amssymb}
\usepackage{amsfonts}
\usepackage{color}
\usepackage{multirow}

\usepackage{algorithmic}
\usepackage{array}
\usepackage{color}
\usepackage{cite}
\usepackage[mathscr]{eucal}
\usepackage{graphics}
\usepackage{float}
\usepackage{cuted}
\usepackage{mathtools}
\usepackage{amssymb}
\usepackage{mathrsfs}
\usepackage{mathdots}
\usepackage{amsmath}

\usepackage[ruled]{algorithm2e}
\usepackage{pgfplots}
\pgfplotsset{compat=1.17}
\usepackage{epstopdf}
\usepackage{tikz}
\usepackage{stfloats}
\usepackage{graphicx}
\usepackage{booktabs}
\usepackage{multirow}
\usepackage{listings}
\usepackage{amsfonts}
\usepackage{bm}
\usepackage{bbm}
\usepackage{tabularx}
\usepackage{makecell}

\usepackage{hyperref}
\usepackage{marvosym}

\usepackage{subfigure}

\title{Evaluating the Influence of Satellite Systems on Terrestrial Networks: Analyzing S-Band Interference}
\author{\IEEEauthorblockN{Lingrui~ZHANG \IEEEauthorrefmark{1}, Zheng~LI \IEEEauthorrefmark{2} and Sheng~YANG\IEEEauthorrefmark{3}}
    \IEEEauthorblockA{
        \IEEEauthorrefmark{1} CentraleSupélec, Orange Innovation, email: lingrui.zhang@student-cs.fr\\
        \IEEEauthorrefmark{2} Orange Innovation, email: zheng1.li@orange.com\\
         \IEEEauthorrefmark{3} CentraleSupélec, email:
sheng.yang@l2s.centralesupelec.fr\\
    }
}
\date{08/07/2024}

\begin{document}
\maketitle

\begin{abstract}
The co-existence of terrestrial and non-terrestrial networks (NTNs) is essential for achieving comprehensive global coverage in sixth-generation cellular networks. Given the escalating demand for spectrum, there is an ongoing global discourse on the feasibility of sharing certain frequencies currently utilized by terrestrial networks (TNs) with NTNs. However, this sharing leads to co-channel interference and subsequent performance degradation. This paper specifically investigates the interference caused by NTNs on TNs in the S-band and its relationship with the relative position between satellite and TN user equipment. We analyzed the transmission mechanisms of satellite signals and employed the ITU two-state model for our interference analysis. Through simulations, we evaluated the interference intensity at different separation distances and slant ranges. Our findings reveal that the angle between the user equipment direction and the sub-satellite point direction from the beam center significantly influences the interference level. Furthermore, we determine the minimum separation distance needed to keep the interference-to-noise ratio of NTN interference below 0 dB.
\end{abstract}

\textbf{Keywords:} Non-terrestrial network (NTN), terrestrial network (TN), low Earth orbit (LEO) satellite, interference, co-existence, data rate, spectrum sharing

\section{Introduction}
Non-terrestrial networks (NTNs) are expected to play a key role in sixth-generation (6G) cellular networks by enhancing coverage and connectivity in remote and underserved areas~\cite{xie2021leo}. NTNs include various networks that operate through the sky, such as satellite networks, high-altitude platform systems, and unmanned aerial vehicles. In recent years, the significant decrease in satellite launch costs and the growing demand for global broadband have led low Earth orbit (LEO) satellites to dominate NTN commercialization, with initiatives such as Starlink, OneWeb and AST SpaceMobile~\cite{lin2021path}.

By integrating with terrestrial networks (TNs), NTNs can address coverage gaps, enhance infrastructure resilience during crises, and support high-bandwidth, low-latency applications like autonomous driving, which are crucial for the 6G era~\cite{azari2022evolution}. However, the deployment of NTN systems faces the challenge of limited spectrum resources. To address the limited bands and the small bandwidths of LEO satellite systems, there is indeed growing discussion about sharing TN bands with NTN systems~\cite{mahboob2024revolutionizing}, which may cause potential co-channel interference for TN user equipments (UEs). 

One such candidate spectrum segment is the S-band (2 GHz to 4 GHz), commonly used by both terrestrial and non-terrestrial systems. Figure \ref{fig:EU S band allocation 2 GHz} shows the allocation of  spectrum to NTNs and TNs at around 2 GHz in Europe.
For instance, terrestrial long-term evolution (LTE) networks often use the 2.5 GHz to 2.7 GHz range. The 2600 MHz band (LTE Band 7 FDD) is specifically dedicated to LTE and LTE Advanced TNs~\cite{TS36101}. Figure \ref{fig:TNs allocation 2.6GHz} shows how the 2600 MHz band is divided between different operators for both uplink and downlink in their LTE networks.
On the other hand, NTNs utilize nearby frequencies, such as 2.0 GHz to 2.2 GHz. In May 2009, Inmarsat and Solaris Mobile were each awarded a 2$\times$15 MHz portion of the S-band by the European Commission, with two years to launch pan-European mobile satellite services  for 18 years. The allocated frequencies are 1.98--2.01 GHz for Earth-to-space and 2.17--2.2 GHz for space-to-Earth communications~\cite{eu_decision_2008,europa_press_release_2009}.

\setcounter{figure}{0}

\begin{figure}[ht]
    \centering
    \includegraphics[width=0.8\linewidth]{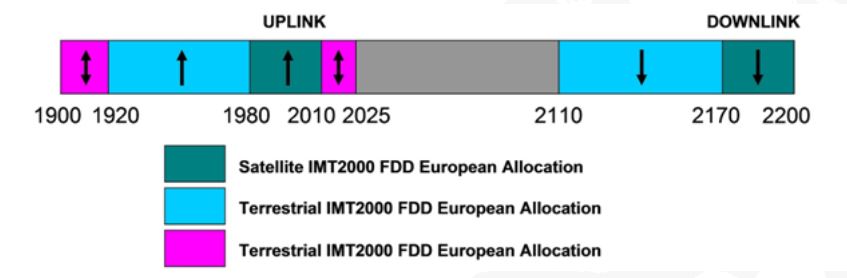}
    \caption{Frequency Allocation around 2 GHz Band in Europe~\cite{ucl_s_band}}
    \label{fig:EU S band allocation 2 GHz}
\end{figure}

\begin{figure}[ht]
    \centering
    \includegraphics[width=0.8\linewidth]{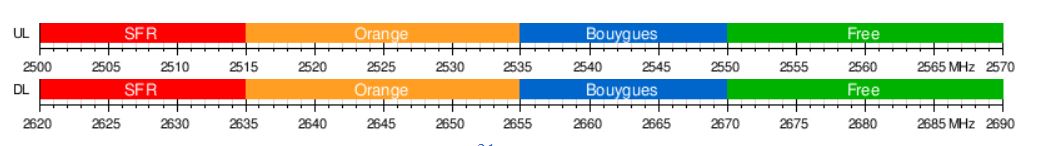}
    \caption{Frequency Allocation of the 2.6 GHz Band in France~\cite{wiki:LTE}}
    \label{fig:TNs allocation 2.6GHz}
\end{figure}

This proximity increases the likelihood of co-channel interference. As a terrestrial operator, understanding and managing interference between terrestrial and satellite networks in shared bands like the S-band is crucial. Therefore, it is essential to investigate the interference strength of a satellite to a TN UE. To quantitatively evaluate the interference caused by NTNs to TN UEs, we adopt the interference-to-noise ratio (INR) as a key metric. 
When the interference power is comparable to or exceeds the noise power (\(\text{INR} > 0\) dB), the NTN interference can significantly degrade the performance of TN UEs~\cite{2022INR}. This highlights the importance of using INR to measure the impact of NTNs on TN systems and to determine acceptable interference levels. In this study, we consider an INR threshold of 0 dB as a critical point; when exceeded, the NTN interference is deemed to cause substantial degradation to TNs performance.

This paper analyzes the interference  power caused by NTNs on TNs in the S-band, ranging from 2 GHz to 4 GHz. This study includes a literature review of satellite signal transmission mechanisms, which provides the foundation for  subsequent simulations and mathematical modeling. Additionally, we apply the ITU two-state model~\cite{ITU-Rreport-P681-11} in generating the channel coefficients for the NTN signal.
Through simulations, we assessed the interference intensity under different relative positions of satellite and TN UEs, considering different slant ranges and separation distances. The separation distance is defined as the shortest distance between the TN UE and the edge of the NTN cell. The results indicate that the angle between the direction of the UE and the sub-satellite point relative to the beam center plays a crucial role in determining the interference levels. 
Furthermore, we determined the minimum separation distance required to keep NTN interference below an acceptable level (\(\text{INR} < 0\) dB) at the TN UE.

\section{System Model}\label{sec:System Model}

\begin{figure}[ht]
    \centering
    \includegraphics[width=0.8\linewidth]{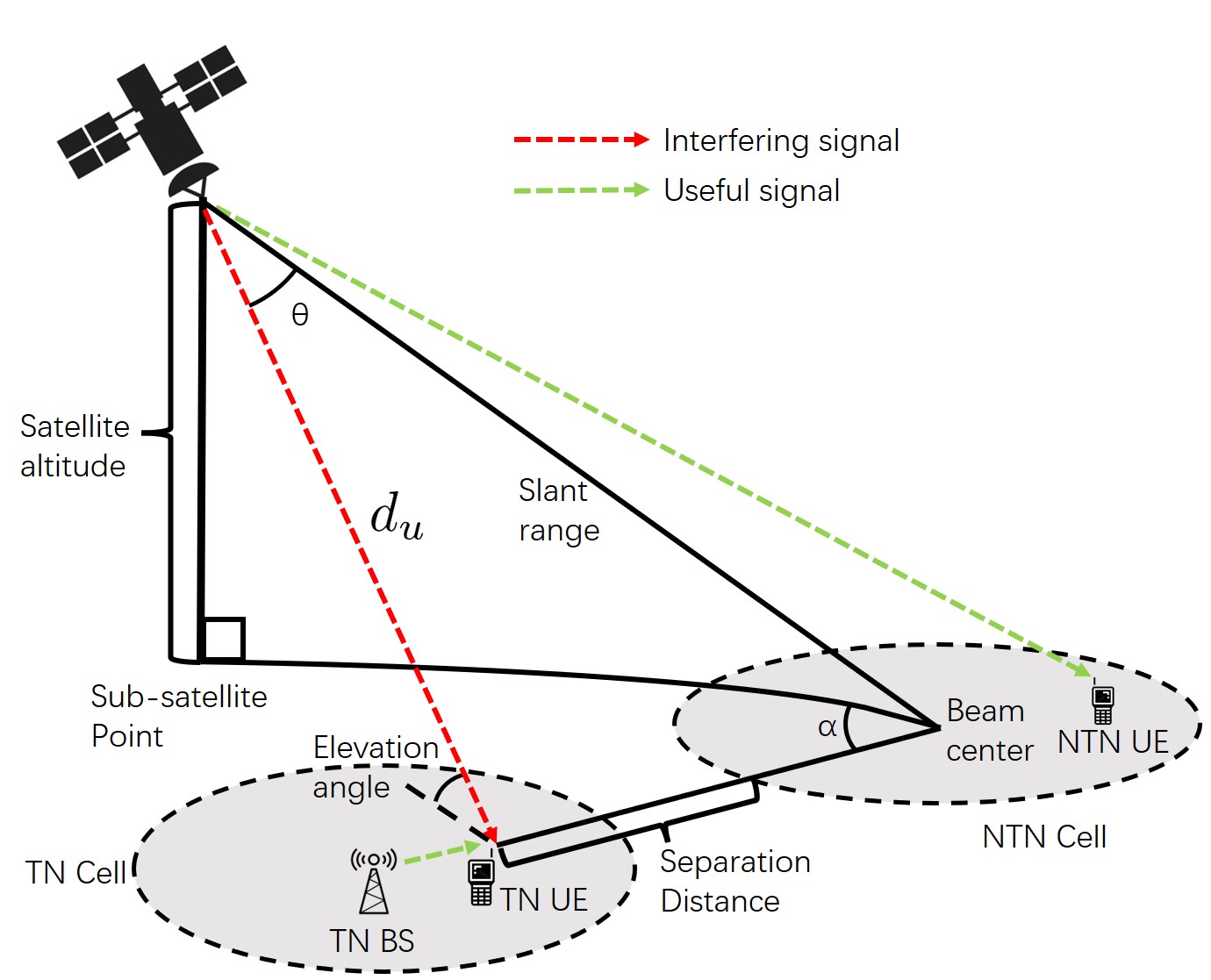}
    \caption{Co-existence scenarios}
    \label{fig:Co-existence scenarios}
\end{figure}

The TN-NTN co-existence case studied in this paper is depicted in Fig. \ref{fig:Co-existence scenarios}. 
This  paper investigates a high throughput satellite (HTS) communication scenario involving a LEO multi-beam satellite sharing the same frequency channels with a TN in the S-band ranging from 2 GHz to 4 GHz.  The LEO satellite, equipped with an array of units, operates in regenerative mode, which enables adaptive payloads and dynamic radio resource management processes. This operational mode significantly enhances the efficiency and flexibility of satellite communications within its designated service area. By leveraging these advanced capabilities,  satellites can better meet the demands of modern communication networks. The connectivity to UE is established through a forward link consisting of two crucial components: the feeder link and the user link. The feeder link plays the vital role of connecting the ground segment's gateway  with the HTS, ensuring seamless communication. On the other hand, the user link establishes a direct connection between the HTS and the respective UEs, enabling seamless interaction through designated sub-channel resources. This study specifically focuses on the downlink  communications within both the NTN and the TN. 

\begin{figure}[!htp]
    \centering
    \includegraphics[width=0.8\linewidth]{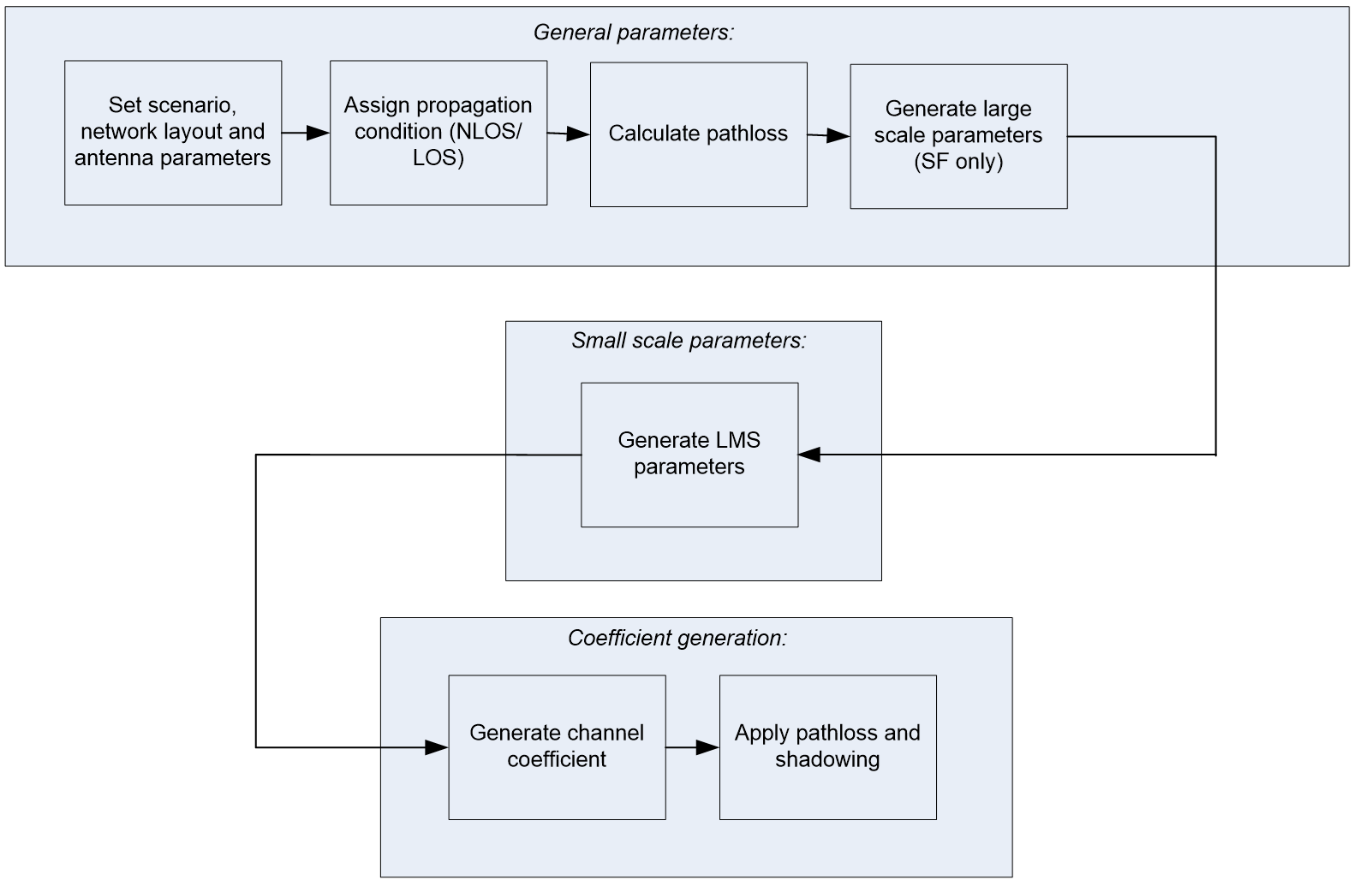}
    \caption{Simplified channel coefficient generation}
    \label{fig:channel_model_generation}
\end{figure}

In the TN, UEs are equipped with an omnidirectional antenna that facilitates connectivity for both the TN and the NTN. The analysis is limited to outdoor conditions, as the building entry loss is substantial, rendering satellite interference negligible for indoor UEs~\cite{TR38811}.
Furthermore, it is assumed that all TN UEs satisfy the flat fading criterion. In this case, the channel coefficients reduce to a single tap, since the channel is not frequency selective. Simultaneously, TNs allow UEs to connect directly to terrestrial base stations without any relay. The process of generating a channel model of interference with flat fading criteria  is depicted in Fig. \ref{fig:channel_model_generation}~\cite{TR38811}.

Owing to the substantial separation—often extending hundreds of kilometers—between the TN UEs and the beam center of the NTN cell, the curvature of the Earth becomes a critical factor, as depicted in Fig. \ref{fig:Co-existence scenarios}.
In this case, the angle between the TN UE direction and the satellite antenna boresight direction is denoted by $\theta$. We call this angle the misalignment angle. The separation distance is determined by subtracting the NTN cell radius from the distance between the NTN cell's beam center and the TN UE. Furthermore, we define $\alpha$ as the angle between the UE direction and the sub-satellite point direction from the beam center. The distance between the UE and the satellite is denoted by $d_u$.

\section{Analysis of Interference Signals via the ITU Two-State Model}

\subsection{Condition of the ITU Two-State Model and Channel Coefficient Generation}
\label{sec: ITU condition}

\begin{figure}[!ht]
    \centering
    \includegraphics[width=0.6\linewidth]{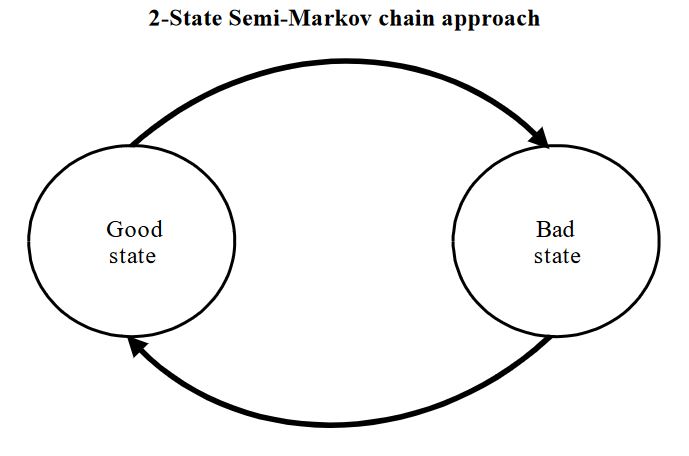}
    \caption{ITU Two-State Semi-Markov chain}
    \label{fig:ITU 2state model}
\end{figure}

\begin{figure}[!ht]
    \centering
    \includegraphics[width=0.95\linewidth]{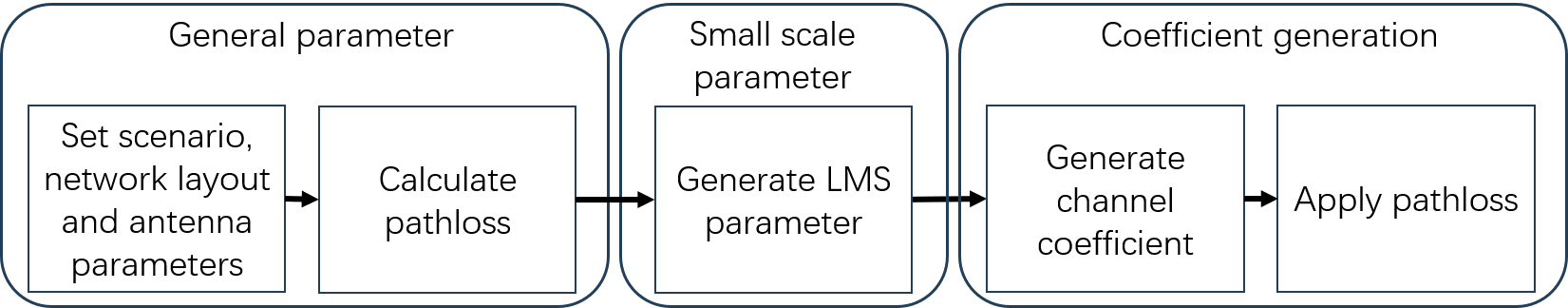}
    \caption{Channel coefficient generation with the ITU two-state model}
    \label{fig:channel_model_generation ITU}
\end{figure}

This paper introduces a methodology for calculating the channel gain of NTN interference signals via the ITU two-state model~\cite{ITU-Rreport-P681-11,TR38811}. 
In the ITU two-state model, the long-term variations of received signal may be described by a semi-Markov chain that including the two distinct states, GOOD and BAD, as shown in Fig. \ref{fig:ITU 2state model}.
To ensure the applicability of the ITU model, several specific conditions must be met. Firstly, the elevation angle should be at least $20^\circ$. Secondly, the frequency range must lie between $1.5$~GHz and $20$~GHz. Additionally, the scenario should demonstrate quasi-line-of-sight (quasi-LOS) conditions, with a fading margin that does not exceed approximately $5$~dB. Furthermore, the channel bandwidth is restricted to $5$~MHz or less. Lastly, the environment must be classified as rural, suburban, or urban.
In this paper, the specific condition in the space model is computed in subsection \ref{sec: Misalignment Angle and Elevation Angle}.
The process of generating a channel model of interference in the ITU two-state model is simplified to Fig. \ref{fig:channel_model_generation ITU}.

\subsection{Satellite Signal Transmission Mechanism with the ITU Two-State Model}

In general, the received power (Rx power) per physical resource block (PRB) 
\(P_{\text{Rx}}\) can be expressed as follows:

\begin{equation}\label{eq:P_Rx}
    P_{\text{Rx}} = P_{\text{TX}} + G(\theta)_{\text{dBi}} + \text{PL} + g + G_{\text{Rx}},
\end{equation}
where $P_{\text{TX}}$  denotes the  power input to the transmitter's antenna, \(G(\theta)_{\text{dBi}}\)  represents the normalized antenna gain, PL indicates the path loss, \(g\) signifies the channel gain affected by small-scale fading, and \(G_{\text{Rx}}\) refers to the antenna gain of the UE~\cite{TR38811}.
The sum of $P_{\text{TX}}$ and $G(\theta)_{\text{dBi}}$ is the effective isotropic radiated power (EIRP) of the satellite.
All the components are expressed in dB or dBi. In this paper, the UE antenna gain is assumed to be $0$~dBi.

\subsection{Satellite Antenna Pattern}
The normalized antenna gain pattern, derived from the theoretical pattern of a circular aperture, serves as a standardized approach for parameterizing Satellite Access Networks (SAN) and conducting related coexistence studies~\cite{TR38811}, despite variations in antenna types. The gain pattern is mathematically represented as follows:
\begin{equation}
    G(\theta)_{dBi}=
    \begin{cases}
        1, & \text{for } \theta = 0^\circ, \\
        4\left| \frac{J_1(ka \sin \theta)}{ka \sin \theta} \right|^2, & \text{for } 0 < |\theta| \leq 90^\circ.
    \end{cases}
\end{equation}
In this equation, \( J_1(x) \) represents the Bessel function of the first kind and first order with argument \( x \). The parameter \( a \) corresponds to the radius of the antenna's circular aperture, while \( k = \frac{2\pi f}{c} \) denotes the wave number. Furthermore, \( f \) denotes the frequency of operation, and \( c \) represents the speed of light in a vacuum. The misalignment angle \( \theta \) is measured from the bore sight of the antenna's main beam. It is noteworthy that \( ka \) is equivalent to the number of wavelengths on the circumference of the aperture and remains consistent regardless of the operating frequency.

\subsection{Path Loss Calculation with the ITU two-state model}
The signal path between a satellite and a terminal undergoes several stages of propagation and attenuation. The path loss is composed of the following components:
\begin{equation}\label{eq:path loss calculation}
    \text{PL}  = \text{PL}_\text{b} + \text{PL}_\text{g} + \text{PL}_\text{c, r}+ \text{PL}_\text{s} + \text{PL}_\text{e},  
\end{equation}
where PL$_\text{b}$ is the basic path loss, $\text{PL}_\text{g}$ is the attenuation due to atmospheric gases, $\text{PL}_\text{c, r}$ is the attenuation due to rain and clouds, $\text{PL}_\text{s}$ is the attenuation due to ionospheric or tropospheric scintillation, and $\text{PL}_\text{e}$ is the building entry loss~\cite{TR38811}. All components are expressed in dB.

In Eq.(\ref{eq:path loss calculation}), rain and cloud attenuation and tropospheric scintillation are considered negligible for frequencies below 6 GHz. Building entry loss is only considered for an indoor station or an indoor UE.
Therefore, rain and cloud attenuation, building entry loss and tropospheric scintillation attenuation are not considered in our simulation. The total path loss is computed as follows: 

\begin{equation}
\text{PL}  = \text{PL}_\text{b} + \text{PL}_\text{g} + \text{PL}_\text{s}.
\end{equation}

\subsubsection{\texorpdfstring{Basic Path Loss \(\text{PL}_\text{b}\)}{Basic Path Loss PL\_b}}

The basic path loss in the dB unit is modeled as
\begin{equation}
    \text{PL}_\text{b} = \text{FSPL}(d, f_c) + \text{SF} + \text{CL},
\end{equation}
where FSPL\( (d, f_c) \) is the free space path loss,  CL  is 
 the clutter loss, and  SF the is shadow fading loss represented by a random number generated by the normal distribution, i.e., \( \text{SF} \sim \mathcal{N}(0, \sigma_{\text{SF}}^2) \). When the UE is in the LOS condition, clutter loss is negligible and should be set to 0 dB in the basic path loss model.

Notably, the ITU two-state model already incorporates clutter loss and shadow fading~\cite{ITU-Rreport-P681-11}. Therefore, the basic path loss in our simulation is calculated as follows:
\begin{equation}
    \text{PL}_\text{b} = \text{FSPL}(d, f_c).
\end{equation}
In MATLAB, we can use the function \(\text{L} = \text{fspl}(R, \text{lambda})\) to calculate \(\text{PL}_\text{b}\).

\subsubsection{Atmospheric Absorption} 
Attenuation by atmospheric gases which is caused entirely by absorption depends mainly on the frequency, elevation angle, altitude above sea level and water vapor density (absolute humidity). At frequencies below 10 GHz, it may normally be neglected. However, for elevation angles below 
 \(10^\circ \), it is recommended that the calculation is performed for any frequency above 1 GHz. Annex 1 of  the Recommendation ITU-R P.676  gives a complete method for calculating gaseous absorption~\cite{ITU-Rreport-P.676-13}. 

In MATLAB, we use the function \(p618PropagationLosses\) to calculate the atmospheric absorption (\(A_g\) in output). 

\subsubsection{Ionospheric Scintillation}

Scintillation corresponds to rapid fluctuations of the received signal amplitude and phase. Ionosphere propagation should be considered for frequencies below 6 GHz.
Specifically, for latitudes between \( \pm 20^\circ \) and \( \pm 60^\circ \) of latitude, \( \text{PL}_\text{s} = 0 \). 
For latitudes above \( \pm 60^\circ \), the presented ITU model is not applicable. However, in these regions, the scintillation phenomena mainly affect the signal phase, with negligible effects on the signal amplitude. Therefore, the choice of \( \text{PL}_\text{s} = 0 \) is also applied for latitudes above \( \pm 60^\circ \).
Finally, for latitudes with a maximum \( \pm 20^\circ \), the \( \text{PL}_\text{s} \) is given by 

\begin{equation}\label{eq: PL_S}
    \begin{split}
        \text{PL}_\text{s}(f_c)&=A_{IS}(f_c)\\
        &=\frac{P_{fluc}(\text{4~GHz})}{\sqrt{2}}\left(\frac{f_c}{4 \ }\right)^{-1.5},
    \end{split}
\end{equation}
where  \(f_c\) is the frequency of  the carrier (GHz) and $P_{fluc}(\text{4~GHz})$ is related to Fig. 6.6.6.1.4-1 in~\cite{TR38811}. This figure provides scintillation occurrence statistics on equatorial ionospheric paths: peak-to-peak amplitude fluctuations, \( P_{\text{fluc}} \) (dB), for 4 GHz reception from satellites in the East at elevation angles of about \( 20^\circ \) (P: solid curves) and in the West at about \( 30^\circ \) elevation (I: dotted curves). The data are given for different times of year and sunspot numbers. From Fig. 6.6.6.1.4-1, the value of $P_{fluc}(\text{4~GHz})$ is \(1.1\) and Eq.\ref{eq: PL_S} is presented by 

\begin{equation}
    \text{PL}_\text{s}(f_c)=0.7778 \cdot \left(\frac{f_c}{4 \ }\right)^{-1.5}.
\end{equation}

\subsection{Generating LMS Parameter and Calculating Channel Coefficient with the ITU Two-State Model}
The channel gain, denoted as \( g \), is calculated via the formula \( g = |h|^2 \), where \( h \) represents the channel coefficient primarily affected by small-scale fading. 
In the ITU two-state model, the signal level is statistically described with a good state (corresponding to LOS and slightly shadowed conditions) and a bad state (corresponding to severely shadowed conditions). The state duration is described by a semi-Markov model. Within each state, fading is described by a Loo distribution, i.e., \( 
\text{Fading} \sim \text{Loo}(M_{A}, \Sigma_{A}, MP)
\). Specifically, the Loo
distribution considers that the received signal is the sum of two components: the direct path signal
and the diffuse multipath. The average direct path amplitude is considered to be normally distributed
and the diffuse multipath component follows a Rayleigh distribution. 
The average LOS power$M_A$ is characterized by a normal distribution $\mathcal{N}(\mu_{M_A },\sigma_{M_A })$. Its correlation of LOS power is presented by $\Sigma_{A}=g_{1 }M_A+g_{2 }$ and average multipath power \(MP\) is computed by \(MP =h_{1 }M_A+h_{2 }\)  , where $\mu_{M_A}$, $\sigma_{M_A}$, $h_{1 }$, $h_{2}$, \(g_1\), \(g_2\) are parameters various for different channels~\cite{ITU-Rreport-P681-11}.

Only outdoor conditions are considered for satellite operations in our simulation since performance requirements are not expected to be met with the available link budget for indoor communications~\cite{TR38811}.
Under flat fading conditions, the LOS probability outlined in section 6.6.1 of~\cite{TR38811} is not used in the ITU two-state model for generating small-scale parameters.

Additionally, the ITU two-state model is preferred due to its ease of simulation in MATLAB with the satellite communication toolbox.

\subsubsection{Methodology}

As discussed in Section \ref{sec:System Model}, the ITU two-state model comprises a good state and a bad state, with their durations governed by a semi-Markov model. Within each state, fading is characterized by a Loo distribution, defined by the following parameters: the mean of the direct signal, the standard deviation of the direct signal, and the mean of the multipath component.
The procedure in section 6.7.1 of~\cite{TR38811} is followed for simulation.

In MATLAB, the "p681LMSChannel system object" can be used to directly obtain channel coefficients  on the basis of scenario parameters such as  the urban environment and elevation angle.

To accurately obtain the channel coefficients, it is essential to configure several specific parameters. First, the carrier frequency must be established, along with the type of propagation environment, which can be categorized as Urban, Suburban, or Rural wooded. Additionally, the elevation angle needs to be specified, with options including \(20^{\circ}, 30^{\circ}, 45^{\circ}, 60^{\circ},\) or \(70^{\circ}\). Furthermore, the Doppler shift correlation coefficient is influenced by various factors, such as the speed of the mobile terminal and the azimuth orientation, which indicates the direction of movement of the ground or mobile terminal. The Doppler shift resulting from satellite movement also plays a crucial role in determining this coefficient.
The value of the Doppler shift due to satellite movement
$f_{d, \text{shift}}$ is computed by 
\begin{equation}
    f_{d, \text{sat}} = \left( \frac{v_{\text{sat}}}{c} \right) \times \left( \frac{R}{R + h} \cos(\alpha_{\text{model})} \right) \times f_c,
\end{equation}
where \( v_{\text{sat}} \) denotes the satellite speed, \( c \) denotes the speed of light, \( R \) denotes the earth radius, \( h \) denotes the satellite altitude, \( \alpha_{\text{model}} \) denotes the satellite elevation angle, and \( f_c \) denotes the carrier frequency.
In our simulation, \( v_{\text{sat}} \) is computed by 
\begin{equation}
     v_{\text{sat}} = \sqrt{\frac{GM}{R + h}}, 
\end{equation}
where \( G \) denotes the gravitational constant and \( M \) is the mass of the Earth. 
The satellite speed, satellite elevation angle and UE speed should be considered constant during the simulation duration if limited by the small number of transmission time intervals (TTIs).

\section{Simulation}
This section presents numerical results to validate the influence of interference signals across different dimensions. 
The parameters for the  simulation are shown in Table \ref{table:satellite parameters}. Notably, the latitude range of \([-20^\circ, 20^\circ]\) significantly influences ionospheric scintillation, making its effect considerable. The EIRP of the satellite is 19.24 dBW according to the parameters in the 6G-NTN project~\cite{6g_ntn_website}.

\begin{table}[ht!]
\caption{Parameters of the satellite}
\centering
\begin{tabularx}{0.7\columnwidth}{|X|c|c|}
\hline
\textbf{Parameter} & \textbf{Unit} & \textbf{S-band} \\
\hline
PRB size & kHz & 180 \\
\hline
Downlink frequency & GHz & 2.17 \\
\hline
Altitude & km & 600 \\
\hline
Antenna aperture (diameter) & m & 0.44 \\
\hline
Maximum antenna gain & dBi & 40.4 \\
\hline
Slant range & km & [600, 1075.19] \\
\hline
Mean slant range & km & 882.38 \\
\hline
NTN Cell Diameter & km & 45 \\
\hline
Latitudes & degree & [-20, 20] \\
\hline
Elevation angle & degree & [20, 90] \\
\hline
Satellite peak EIRP per PRB & dBW & 19.24 \\
\hline
Earth radius & km & 6378 \\
\hline
Equivalent Temperature & K & 2303.55\\
\hline
TN Average AWGN power per PRB & dB & -112.39\\
\hline
Channel gain  & dB & 1.2 \\
\hline
\end{tabularx}

\label{table:satellite parameters}
\end{table}

In subsection \ref{sec: Misalignment Angle and Elevation Angle}, we examine the range of misalignment and elevation angles within our space model. Additionally, we compute the specific separation distance that satisfies the conditions specified in the ITU two-state model, as described in subsection \ref{sec: ITU condition}.
Subsection \ref{sec:max channel gain} focuses on calculating the average channel gain across different environments and elevation angles, presenting results with a 95\% confidence interval. Table \ref{table: scenario} outlines the environmental parameters and elevation angles for scenarios where statistical data are available for the two-state model, as referenced in~\cite{ITU-Rreport-P681-11}. For the simulations, we consider Urban, Suburban, and Rural wooded environments.

In subsections \ref{sec:SR} and \ref{sec:SD}, specific UE parameters, including environment and terminal speed, are not considered. Instead, the analysis emphasizes the system's worst-case scenario, leveraging the maximum channel gain computed earlier.
Subsection \ref{sec:SR} investigates a scenario where a user, positioned at a fixed distance from the terrestrial network, attempts to establish a connection with the satellite network. This evaluation assesses how the satellite's relative position influences additional interference, quantified as INR, experienced by adjacent TN users.
Subsection \ref{sec:SD} examines the additional interference experienced by TN UEs at various separation distances from a NTN, considering the satellite network's design parameters. Furthermore, it estimates the number of TN UEs beyond the specified separation distance that encounter acceptable levels of interference, defined as INR below 0 dB.

\begin{table}[ht!]
\centering
\caption{Available elevation angle in different environment for frequencies between 1.5 and 3~GHz}
\begin{tabular}{|c|l|}
\hline
\textbf{Environment} & \textbf{Available Elevation Angles (degrees)} \\
\hline
Urban & 20, 30, 45, 60, 70 \\
\hline
Suburban & 20, 30, 45, 60, 70 \\
\hline
Village & 20, 30, 45, 60, 70 \\
\hline
Rural wooded & 20, 30, 45, 60, 70 \\
\hline
Residential & 20, 30,  60, 70 \\
\hline
\end{tabular}

\label{table: scenario}
\end{table}

\subsection{Range of Misalignment Angle and Elevation Angle}\label{sec: Misalignment Angle and Elevation Angle}
\begin{figure}[!htp]
    \centering
    \includegraphics[width=0.8\linewidth]{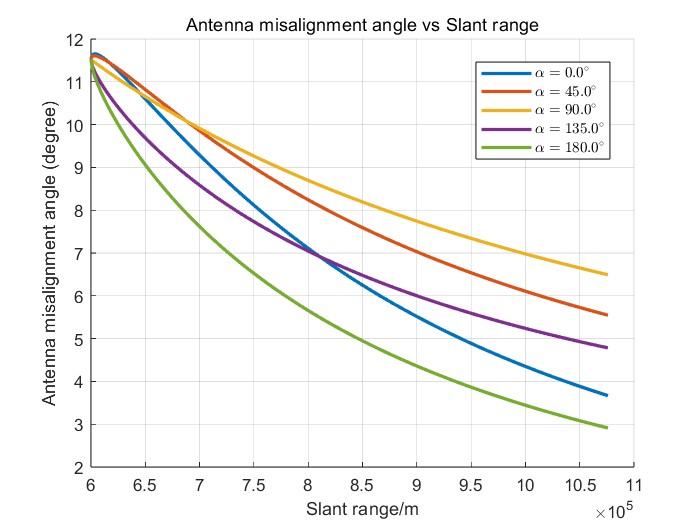}
    \caption{Relationships between the misalignment angle and slant range (separation distance= 100 km)}
    \label{fig:misalignmentAngle vs slant range}
\end{figure}

\begin{figure}[!htp]
    \centering
    \includegraphics[width=0.8\linewidth]{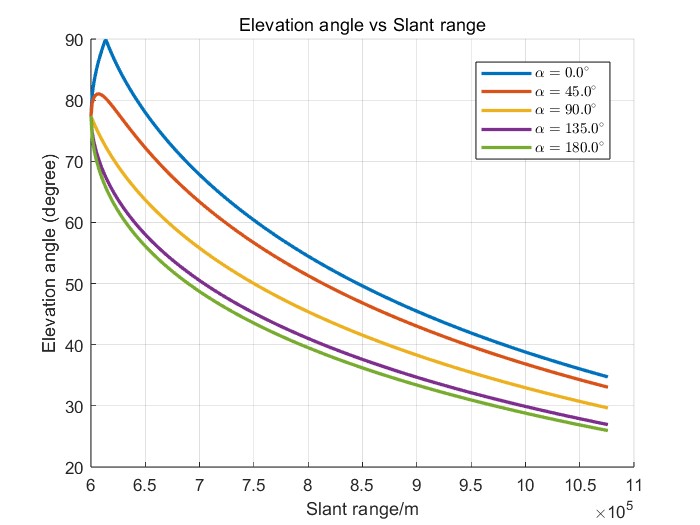}
    \caption{Relationships between the elevation angle and slant range (separation distance= 100 km)}
    \label{fig:elevationAngle vs slant range}
\end{figure}

\Cref{fig:misalignmentAngle vs slant range,fig:elevationAngle vs slant range} illustrate the relationship between the slant range and parameters such as misalignment angle and elevation angle, given a fixed separation distance of 100 km.
Fig \ref{fig:misalignmentAngle vs slant range} shows how the antenna misalignment angle changes with slant range for different angles $\alpha$, given a fixed separation distance of 100 km. 
The data indicate that as the slant range increases, the misalignment angle decreases in all directions. 
However, the descent rate varies for different misalignment angles, being fastest at $\alpha=180^\circ$ and slowest at $\alpha=90^\circ$. 
Figure \ref{fig:elevationAngle vs slant range} shows the correlation between the elevation angle and the slant range for various $\alpha$, given a fixed separation distance of 100 km. It depicts how changes in slant range affect the elevation angle for different $\alpha$ values. Notably, the blue line representing $\alpha = 0^\circ$ represents the highest elevation angles, whereas higher $\alpha$ values correspond to lower elevation angles at the same slant range. This decreasing trend in elevation angle is more pronounced at shorter slant ranges and gradually moderates as the slant range extends.  

\begin{figure}[!htp]
    \centering
    \includegraphics[width=0.8\linewidth]{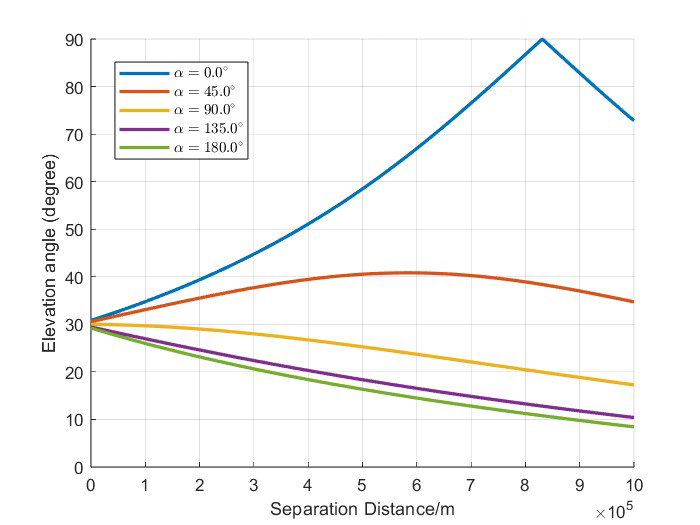}
    \caption{Relationships between the elevation angle and separation distance (slant range= 1075km)}
    \label{fig:elevation angle vs seraparion distance SR1075}
\end{figure}

To ensure a comprehensive understanding of the elevation angle’s behavior across various
separation distances, we will examine the elevation angle at the maximum slant range of 1075
km. This approach is based on the established observation that the elevation angle decreases as
the slant range increases, a characteristic that remains consistent across different separation distances. By focusing on the maximum slant range, we aim to determine the minimum elevation
angle of the system, thereby providing critical insights into its operational parameters. Figure \ref{fig:elevation angle vs seraparion distance SR1075}
demonstrates that the elevation angle varies significantly with the separation distance and the
beam alignment angle $\alpha$. The elevation angle peaks at a mid-range distance for $\alpha$ = 0.0° and
shows a continuous increase for $\alpha$ = 45.0°. In contrast, for angles of 90.0°, 135.0°, and 180.0°,
the elevation angle generally decreases with increasing separation distance. Furthermore, the
minimum of elevation angle in Fig. \ref{fig:elevation angle vs seraparion distance SR1075} is 8° which is also the minimal elevation angle in our
space model.  

Not all UEs at any separation distance or slant range meet the accessible elevation angle specified in Table \ref{table:satellite parameters}. Table \ref{tab:max SD vs SR} shows the maximum separation distance for the ITU two-state model at key slant ranges.

\begin{table}[ht!]
    \centering
\caption{Maximum separation distance in different slant range}
\label{tab:max SD vs SR}
    \begin{tabular}{|c|c|} \hline 
         Slant range / km& Maximum separation distance / km\\ \hline 
         600& 1150\\ \hline 
         882.38& 550\\ \hline 
         1075& 320\\ \hline
    \end{tabular}

\end{table}

\subsection{Max Channel Gain of Interference Signal}\label{sec:max channel gain}
Determining the value of channel gains in a  simulation presents significant challenges, specially owing to the difficulty in accurately setting the parameters of UEs, such as their environmental context and movement rate. Given that our primary concern is the maximum signal interference, minimizing fading is crucial.

\begin{figure}[!htp]
    \centering
    \includegraphics[width=0.8\linewidth]{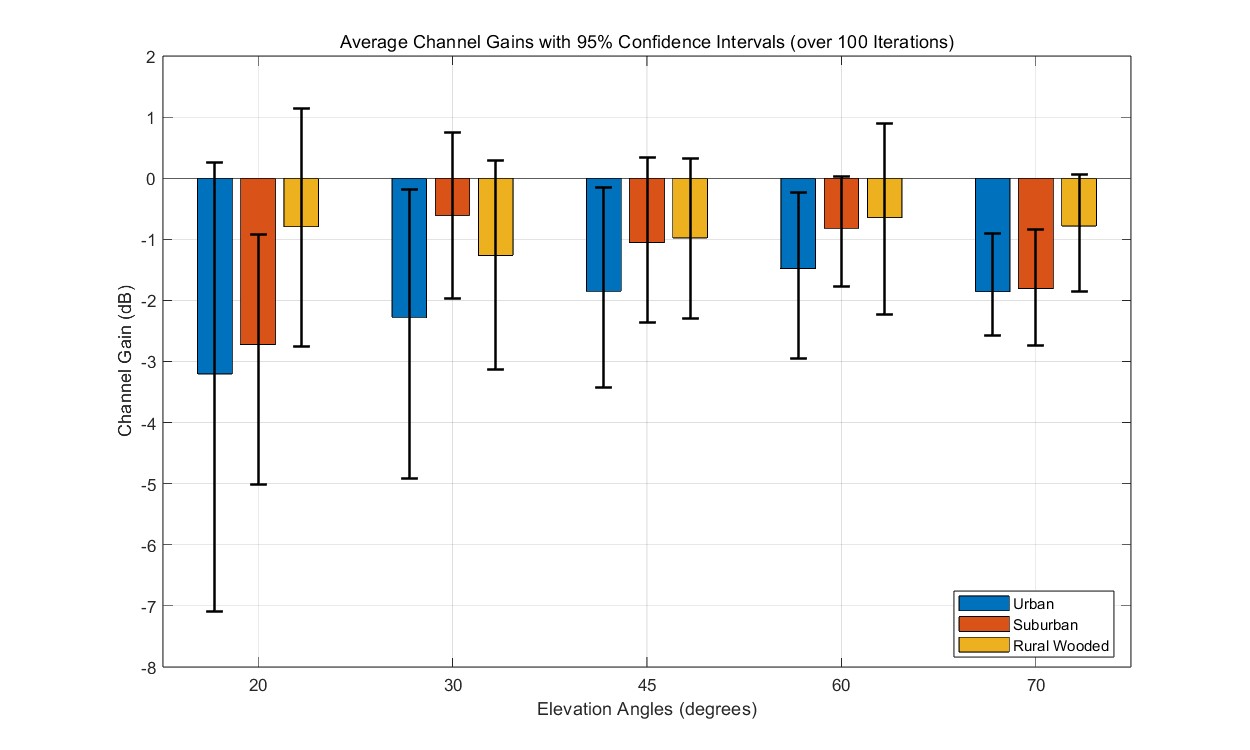}
    \caption{Average mean channel gains under different environments and elevation angles}
    \label{fig:ChannelGains_VS_EnvironmentsAndElevationAngles}
\end{figure}

To accurately calculate the strength of the interfering signal, it is essential to consider the attenuation value under optimal conditions. The average channel gains, along with their 95\% confidence intervals, are depicted in Fig. \ref{fig:ChannelGains_VS_EnvironmentsAndElevationAngles} for various elevation angles (20°, 30°, 45°, 60°, and 70°) across Urban, Suburban, and Rural Wooded environments. 
A key observation is that the maximum value of all confidence intervals consistently stays below 1.2 dB, indicating that the channel gain in most scenarios is limited by this upper bound.
Furthermore, the average channel gains are predominantly negative or close to zero, highlighting the overall attenuation present in these conditions. The confidence intervals exhibit significant variation, with broader intervals at lower elevation angles (e.g., 20° and 30°), likely due to increased multipath propagation or environmental interference. Conversely, narrower intervals at higher angles (e.g., 60° and 70°) suggest more stable signal performance. Given that no scenario exceeds the 1.2 dB threshold, this value can be conservatively adopted as the maximum channel gain for interference signal modeling in subsequent experiments. This approach enables effective modeling and analysis of the impact of extreme conditions on signal interference.

\subsection{Results across various Slant Ranges}\label{sec:SR}

\begin{figure}[!htp]
    \centering
    \includegraphics[width=0.8\linewidth]{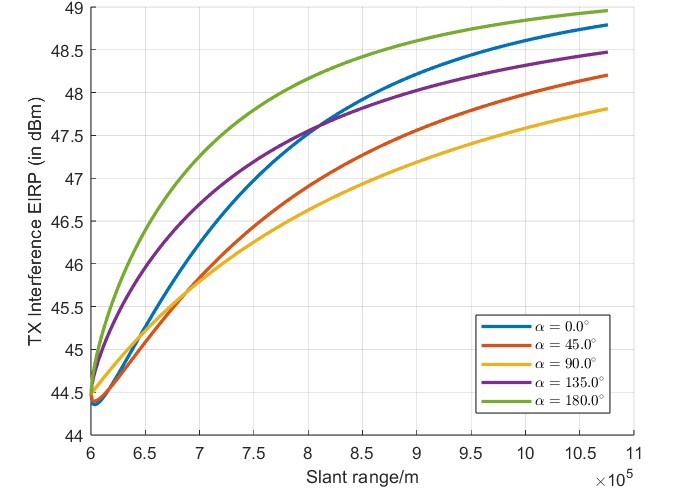}
    \caption{TX Interference EIRP for different slant ranges}
    \label{fig:TX EIRP}
\end{figure}

\begin{figure}[!htp]
    \centering
    \includegraphics[width=0.8\linewidth]{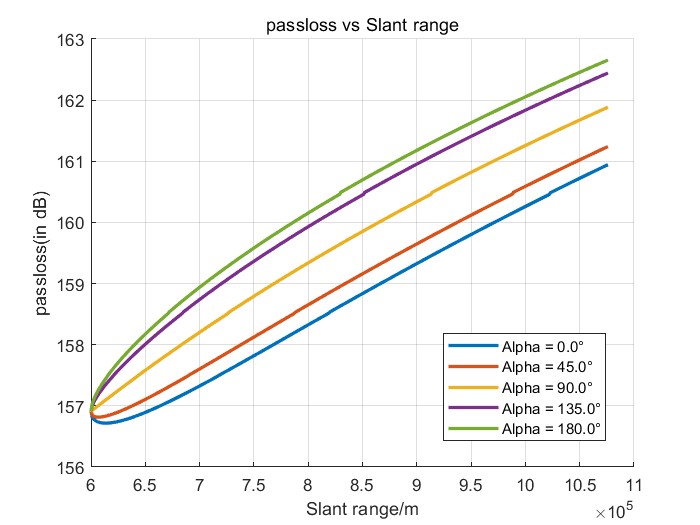}
    \caption{Path loss  at various slant ranges}
    \label{fig:path loss}
\end{figure}

\begin{figure}[!htp]
    \centering
    \includegraphics[width=0.8\linewidth]{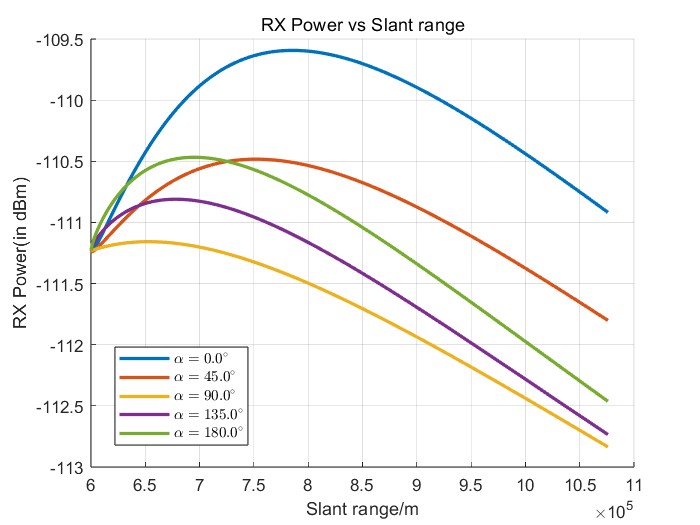}
    \caption{Received interference power per PRB on various slant ranges}
    \label{fig:RX_power}
\end{figure}

\begin{figure}[!htp]
    \centering
    \includegraphics[width=0.8\linewidth]{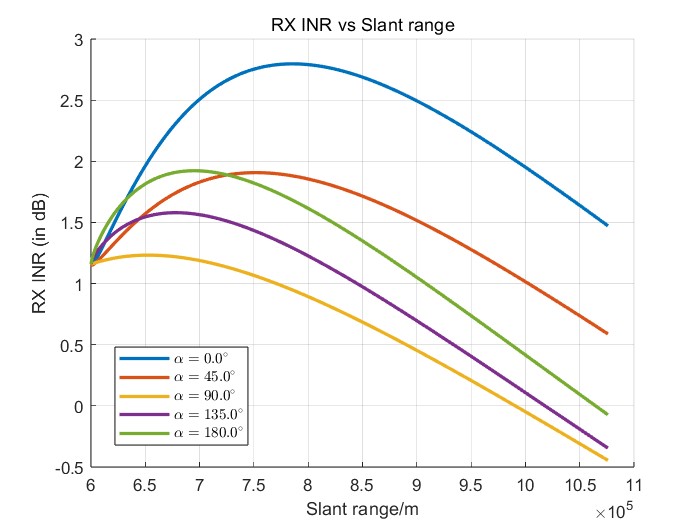}
    \caption{RX INR for various slant ranges}
    \label{fig:RX INR SR}
\end{figure}

This subsection presents an interference analysis conducted to assess the impact of interference signals to TN UEs with a fixed separation distance.  
\Cref{fig:TX EIRP,fig:path loss,fig:RX_power,fig:RX INR SR}  show the power on the transmitter side, the path loss and the received power on various slant ranges. Figure \ref{fig:RX_power} depicts the relationship between the received interference power and slant range for various angles \(\alpha\). Each curve except \(\alpha=90^\circ\) shows that the interference power initially increases with the slant range, reaches a peak, and then decreases. This is because  the interference EIRP grows faster than the path loss in the first half, as shown in Fig. \ref{fig:TX EIRP}, while in the second half the path loss is more dominant. Moreover, lower angles (\(\alpha = 0^\circ\) and \(\alpha = 45^\circ\)) result in higher peaks, indicating a lower path loss. Conversely, higher angles (\(\alpha = 135^\circ\) and \(\alpha = 180^\circ\)) result in lower peaks and earlier occurrences due to increased path loss, as shown in Fig. \ref{fig:path loss}, and  the fast decay increasing rate of antenna gain. This trend highlights the significant impact of the angle on interference power levels in satellite communication systems.

\Cref{fig:RX INR SR} illustrates the TN INR as a function of slant range across various azimuth angles (\(\alpha\)). It is noteworthy that, at substantial slant ranges exceeding \(100\) km, certain azimuth angles, specifically \(\alpha = 0^\circ\) and \(\alpha = 45^\circ\), display RX INR values surpassing 0 dB. This indicates unacceptable levels of interference, underscoring the potential for satellites positioned in these specific directions to cause persistent interference, even at considerable distances. Therefore, terrestrial operators should not open their TNs' frequency bands to satellites unless effective interference mitigation strategies are developed and implemented.

\subsection{Result across various Separation Distance}\label{sec:SD}

This subsection presents an interference analysis conducted to assess the impact of interference signals across various separation distances. In  the first four experiments, the slant range is fixed at $882.38$ km with separation distances ranging from 0 to 550 km which ensures an elevation angle of at least $20^\circ$. 
We then calculate the minimal separation distance at which the INR is less than 0 dB for different slant ranges. 

\begin{figure}[!htp]
    \centering
    \includegraphics[width=0.8\linewidth]{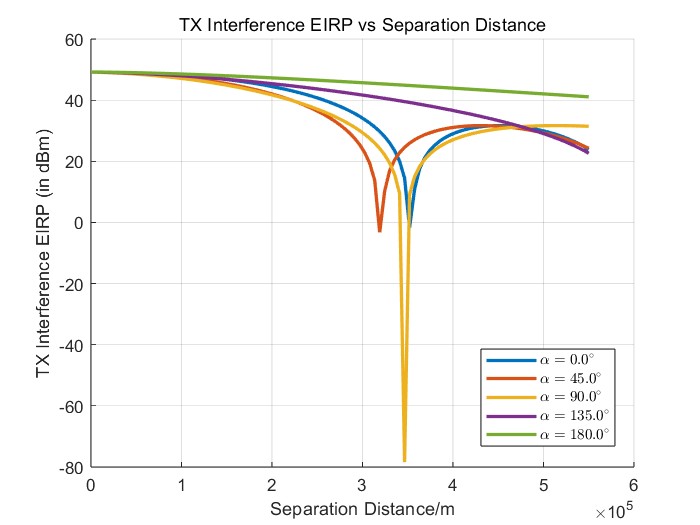}
    \caption{TX EIRP for various separation distances}
    \label{fig:TX EIRP SD}
\end{figure}
\begin{figure}[!htp]
    \centering
    \includegraphics[width=0.8\linewidth]{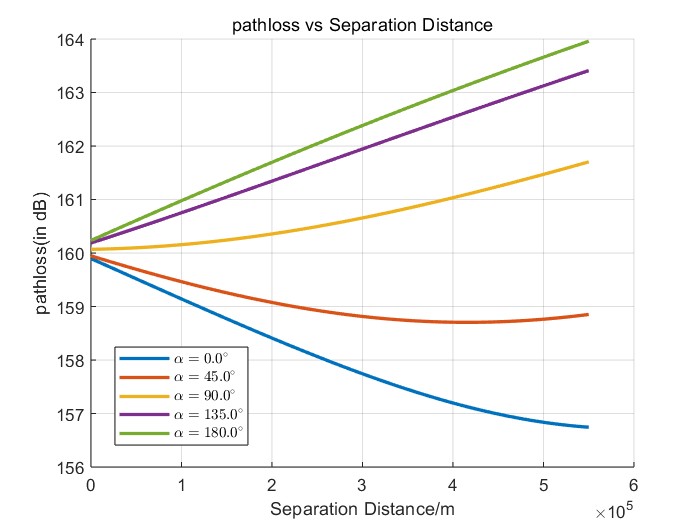}
    \caption{Path loss for various separation distances}
    \label{fig:pathloss SD}
\end{figure}
\begin{figure}[!htp]
    \centering
    \includegraphics[width=0.8\linewidth]{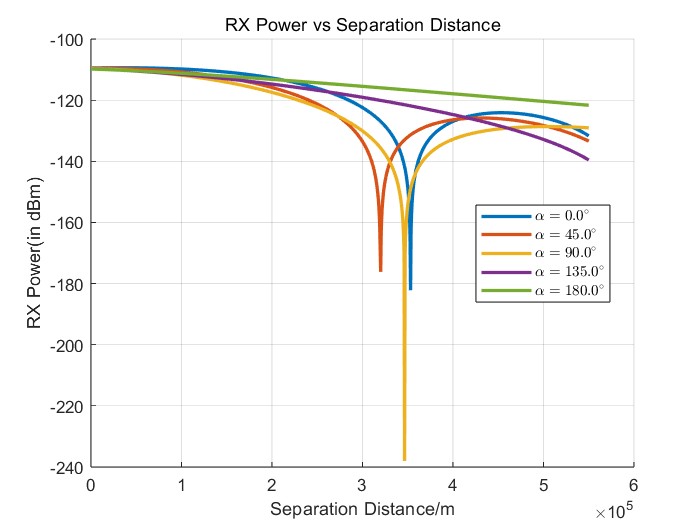}
    \caption{RX power for various separation distances}
    \label{fig:RX power SD}
\end{figure}
\begin{figure}[!htp]
    \centering
    \includegraphics[width=0.8\linewidth]{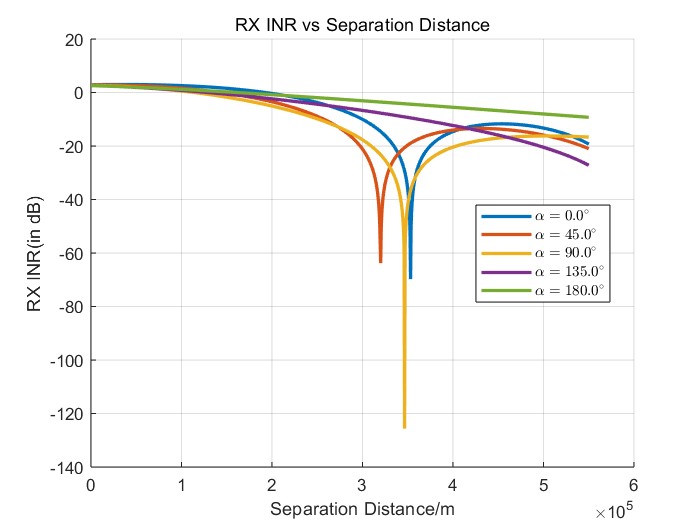}
    \caption{RX INR for various separation distances }
    \label{fig:RX INR SD}
\end{figure}

\Cref{fig:TX EIRP SD,fig:pathloss SD,fig:RX power SD} depict the relationships between  the separation distance and several metrics, including the RX power,  the TX interference EIRP, and path loss, at various angles ($\alpha$). As the separation distance increases, the EIRP generally decreases. 
Significant drops occur around $320$ km for angles of $0^\circ$, $45^\circ$, and $90^\circ$, suggesting that at these distances, the orientation of UEs aligns with the null point of the satellite's radiation pattern.
In contrast, the curves for $135^\circ$ and $180^\circ$ show only slight decreases, as the misalignment angle from the satellite is smaller at the same separation distance than at other angles. Consequently, neither angle reaches the null point, even at a separation distance of 550 km.
The path loss plot generally shows an increase with separation distance at most angles. However, at angles of $0^\circ$ and $45^\circ$, the behavior differs. Initially, the path loss decreases before increasing again as the distance increases. This phenomenon arises because the TN UE moves closer to the satellite as the separation distance increases at these angles.
Furthermore, the trend of each curve in Fig. \ref{fig:RX power SD} and Fig. \ref{fig:RX INR SD} closely resembles that of the EIRP in Fig. \ref{fig:TX EIRP SD}, indicating that the EIRP is the primary factor affecting variations in the RX power.

\begin{figure}[!htp]
    \centering
    \includegraphics[width=0.8\linewidth]{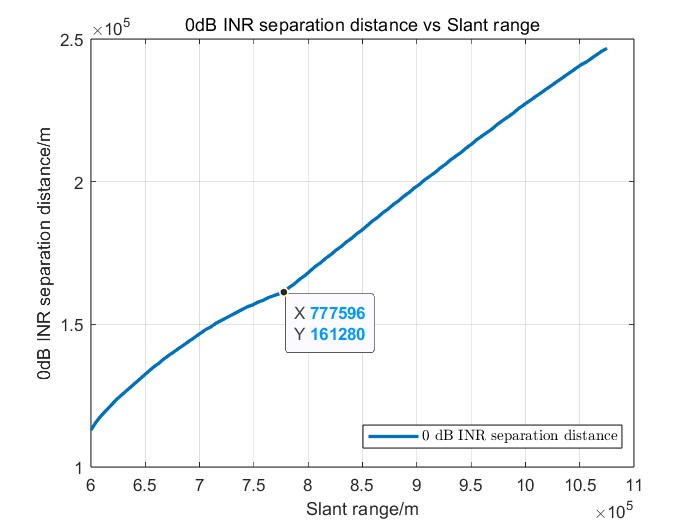}
    \caption{0 dB separation distance for various slant ranges}
    \label{fig:0 dB separation distance}
\end{figure}

\begin{figure}[!htp]
    \centering
    \includegraphics[width=0.8\linewidth]{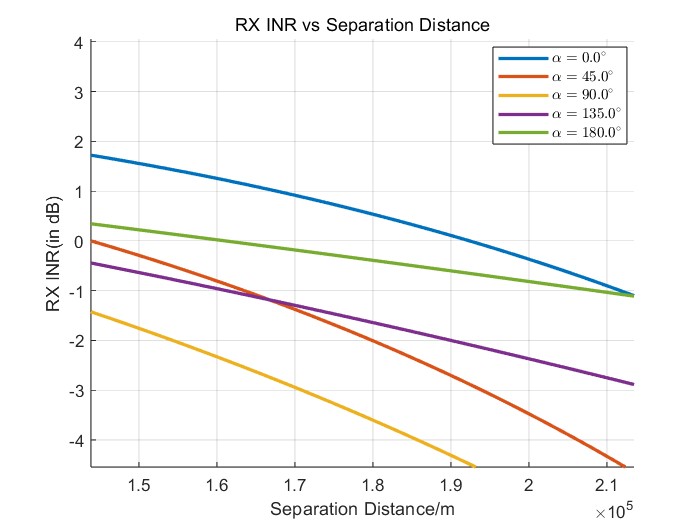}
    \caption{RX INR vs. Separation Distance for Slant Ranges at 883 km}
    \label{fig:RX_INR_SD_883km}
\end{figure}

\begin{figure}[!htp]
    \centering
    \includegraphics[width=0.8\linewidth]{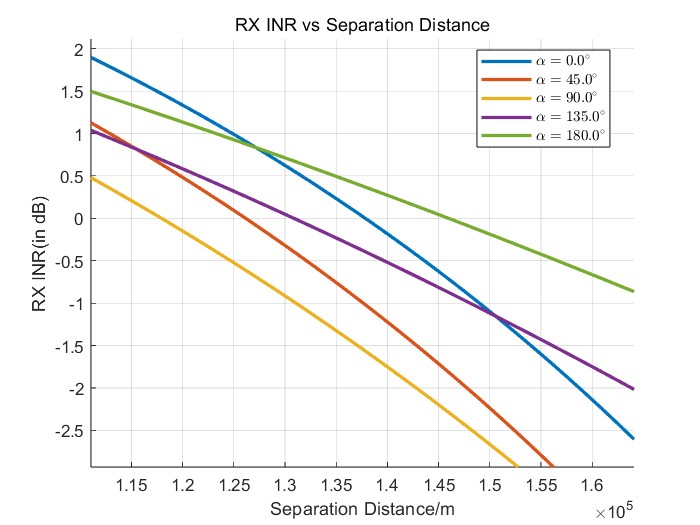}
    \caption{RX INR vs. Separation Distance for Slant Ranges at 700 km}
    \label{fig:RX_INR_SD_700km}
\end{figure}

Figure \ref{fig:0 dB separation distance} illustrates the required separation distance necessary to achieve an acceptable RX INR of 0 dB as a function of slant range. A notable shift in the trend of separation distance occurs around 770 km. For slant ranges between 600 km and 770 km, the maximum separation distance is primarily influenced by satellites positioned at \(\alpha = 180^\circ\). This is evident in Fig. \ref{fig:RX_INR_SD_700km}, where the INR at this angle \(\alpha\) is predominant. Conversely, for slant ranges exceeding 770 km and extending up to 1075 km, the largest required separation distance is associated with satellites at \(\alpha = 0^\circ\), as demonstrated in Fig. \ref{fig:RX_INR_SD_883km}. This transition underscores the directionality of the interference, indicating that the direction of the satellites causing the most interference to the TN UEs changes significantly as the separation distance varies. 
Additionally, all separation distances are within 320 km, confirming the applicability of the ITU two-state model for all scenarios in this analysis. 
These results provide valuable insights for terrestrial operators in assessing the extent to which their TNs are susceptible to interference from a specific NTN. By analyzing the required separation distances, operators can identify the geographic scope within which TNs are significantly affected by interference. This understanding is critical for making informed decisions about whether to share TN frequency bands with NTNs and for devising effective strategies to mitigate interference in shared spectrum environments.

\section{Conclusion and Future Work}

In this paper, we investigate satellite transmission mechanisms and analyze variations in the interference intensity across different slant ranges and separation distances. Our findings reveal that the angle between the UE direction and the sub-satellite point direction significantly affects the RX power, with notable differences observed. By applying the ITU two-state model in interference analysis, we identify the EIRP as a primary factor influencing  the RX power changes and compute the minimal 0 dB separation distance for various slant ranges. Future research will explore more complex scenarios, including multi-user, multi-base station, and multi-satellite configurations, to gain deeper insights.


\bibliographystyle{IEEEtran}
\bibliography{sample}

\end{document}